\documentstyle [12pt,epsf]{article}

\input{epsf}
\begin{document}
\begin{titlepage}
\begin{center}

\vspace{-0.7in}

{\large \bf Finite Size Effects in Thermal Field Theory}\\
\vspace{.3in}{\large\em N.F.Svaiter\,\,\footnotemark[1]}\\ 

Centro Brasileiro de Pesquisas Fisicas-CBPF\\ Rua Dr. Xavier Sigaud 150,\\
Rio de Janeiro, RJ, 22290-180, Brazil\\

\subsection*{\\Abstract}
\end{center}

We consider a neutral self-interacting massive scalar 
field defined in a d-dimensional Euclidean space.  
Assuming thermal equilibrium,
we discuss the one-loop perturbative renormalization 
of this theory in the presence of rigid boundary surfaces 
(two parallel hyperplanes),
which break translational symmetry. 
In order to identify the singular parts of the one-loop 
two-point and four-point 
Schwinger functions, we use a combination of dimensional and zeta-function 
analytic regularization procedures. 
The infinities which occur in both the 
regularized one-loop two-point and four-point 
Schwinger functions fall into two distinct classes: 
local divergences that could be renormalized 
with the introduction of the usual bulk counterterms, and surface  
divergences that demand countertems concentrated on the boundaries.
We present the detailed form of the surface divergences and
discuss different strategies that one can assume 
to solve the problem of the surface divergences. 
We also briefly mention how to 
overcome the difficulties generated by infrared divergences
in the case of Neumann-Neumann boundary conditions.

\footnotetext[1]{e-mail:nfuxsvai@cbpf.br}

PACS numbers: 03.70+k,04.62.+v.

\end{titlepage}
\newpage \baselineskip .37in

\section{Introduction}

The Casimir effect is the manifestation of the 
zero-point energy of the quantized electromagnetic field, in the 
presence of metallic plates  \cite{casi}. A very simple calculation  
predictes that in a four-dimensional 
spacetime, uncharged perfectly conducting parallel
plates should attract each other with a force per unit area 
$F(L)\propto\frac{1}{L^{4}}$,
where $L$ is the distance between the plates.  
Extensive reviews of this subject can be 
found in Refs. \cite{pmg} \cite{mer} \cite{livro} \cite{mb} 
\cite{milton}. As stressed by Milloni et al. \cite{milloni},
a brief argument showing that the zero point-energy 
associated with the quantized electromagnetic field
must have a physical meaning was 
already given by Einstein and Stern \cite{einstein}. These authors 
noted that a zero-point energy seems necessary in order to 
avoid a 
first-order quantum correction to $\beta^{-1}$ in the classical limit $\beta\,>> \omega$ in  
Planck's expression for the average energy of an oscillator in equilibrium 
with radiation at temperature $\beta^{-1}$.

Although the vacuum energies of different 
physical configurations are formally divergent, their diference 
can be finite. In the 
case of a free scalar field, 
interacting only with boundary surfaces,
the Casimir 
approach can be summarized as follows: 
first a complete set of modes solutions 
of the Klein-Gordon equation satisfying 
appropriate boundary conditions, and their  
respective eigenfrequencies are presented.
Next, the divergent zero-point energy is 
regularized by the introduction of an ultraviolet cut-off. Finally, the 
polar part of the regularized energy is removed using a renormalization 
procedure. This procedure was first discussed by Fierz \cite{fierz}
a long time ago, followed by Boyer \cite{boyer} 
and also by Svaiter and Svaiter \cite{nfs1} \cite{nfs2}. 
In these two last references,
an attempt to clarify the relation between the cut-off 
method and analytic regularization procedures in Casimir effect
has been developed. 
In particular, 
in these papers an analytic regularization procedure was 
interpreted as a cut-off method, and using a mixed cut-off in the 
regularized zero-point energy, it was possible to unify these
two methods both in two- and three-dimensional spacetimes.
Further, a general proof was given that when 
the introduction of an exponential cut-off yields
an analytic function with a pole at the origin, then the analytic 
regularization using the zeta function (or a generalization for the zeta function) is equivalent to the aplication of a cut-off with 
the subtraction of the singular part at the origin
\cite{provag} \cite{provag2}. 
More recently, Fulling offered an interesting discussion 
with regard to the problems in the renormalization program 
devised to find 
the renormalized vacuum stress-tensor in different field theories \cite{fulling}.

It is important to point out that these results are valid 
at one-loop 
level and one is dealing with free fields only. It is 
clear that the formalism must be generalized to 
take into account the case of self-interacting fields.   
Although higher-loop corrections 
to the Casimir effect seem beyond  
experimental reach today, theoretically such corrections 
are certainly of interest. Nevertheless, with the  
exception for some few papers, 
only global issues have been discussed 
in the study of radiative corrections to the Casimir effect.
One such exception is the discussion presented by 
Robaschik et al. \cite{rsw}. With this scenario in mind, 
it is natural to ask the important question: how to implement 
the perturbative renormalization algorithm,   
assuming the presence of rigid boundaries (hard-walls),
using the standard weak-coupling perturbative expansion in quantum 
field theory, that is, how to 
implement the one-loop perturbative renormalization of a 
self-interacting scalar theory, assuming boundary 
conditions which do break translational symmetry. Our 
aim when studying these
issues is linked to the following question: does the infrared 
problem have a solution in theories where translational 
invariance is broken? Note that temperature effects can solve the 
infrared problem in some models in quantum field 
theory \cite{linde}; for a recent treatment in non-abelian gauge theories at 
high temperature, and the infrared problem, 
see for example Ref. \cite{braaten}. 
Also, in massless scalar $\lambda\varphi^{4}$ theory, 
if  thermal equilibrium with a reservoir
is assumed, the infrared problem 
can be solved after a ressumation procedure. The standard is
to use the Dyson-Schwinger equation to write a non-perturbative 
version of the self-energy gap equation, or to use the composite operator
formalism \cite{kapusta} \cite{re2} \cite{ananos}.

We would like to call the attention of the reader that 
there are some disagreements in the literature as to implementing 
the one-loop perturbative renormalization in finite size systems 
when translational invariance is broken.
In the one-loop approximation, Albuquerque et al. \cite{erro1}
found that the mass counterterm depends
on the size of the compact dimension in the  $\lambda\varphi^{4}$ theory. 
Also, Malbouisson et al. \cite{erro2} assumed a 
self-interacting scalar field confined between two infinite parallel plates,
and using the techniques developed by Ananos et al. \cite{ananos} these 
authors didn't find any 
surface countertem in the $\lambda\varphi^{4}$ theory at finite 
temperature. Furthermore, they were able to define temperature and  
size-dependent mass and coupling constant terms in systems where
translational invariance is broken.

The purpose of this paper 
is to present a detailed calculation of the one-loop renormalization 
of the 
$\lambda\varphi^{4}$ theory at finite temperature, 
assuming that one of the 
spatial coordinates is confined to a finite interval.
Since this assumption is not sufficient to explicitly breaking 
the translational symmetry, we will further introduce boundary 
surfaces where the field satisfies appropriate boundary 
conditions. In this situation, the breaking of the translational invariance
of the theory is ensured. This paper is a natural continuation  
of the papers of Fosco and Svaiter \cite{fosco} and also Caicedo and Svaiter \cite{caicedo}. 
Our aim is to further the understanding of the renormalization procedure in 
systems at finite temperature where there is a break of translational 
symmetry. We will discuss the Dirichlet-Dirichlet (DD) and also the
Neumann-Neumann (NN) boundary conditions. For the 
Dirichlet-Dirichlet boundary 
conditions, the model is free of infrared divergences. In the 
Neumann-Neumann boundary conditions case, infrared divergences 
associated with zero modes will appear for bare massless fields.
We show that there is no clear meaning for a thermal- or size-dependent
mass in such situations. Consequently, can not be used 
to solve  
the infrared problem in the case of Neumann-Neumann boundary conditions a resummation procedure.

The organization of the paper is the following: 
in the section II we sketch the
general formalism of the theory, deriving the one-loop two-point and  
four-point functions. In section
III we use two different analytic regularization 
procedures, i.e, dimensional regularization and zeta-function 
analytic regularization, to identify the 
polar contributions that appear in the expressions of the 
one-loop two-point and four-point 
Schwinger functions. In section IV we renormalize 
the four-point 
Schwinger function and the problem 
for the infrared divergences is raised.
In the conclusions we will   
discuss alternative solutions for the problem of the surfaces counterterms. 
In this paper we use $\hbar=c=k_{B}=1$.

\section{General Formalism and the Finite Temperature Generating 
Functional of Schwinger Functions}

The static properties of finite temperature 
field theory can be derived from the partition function \cite{justin}. 
To obtain the partition function the 
starting point is the Feynman, Matheus and Salam approach 
\cite{matheus}. Thus, let us consider 
the generating functional of (complete) Green's functions 
for a self-interacting scalar field theory defined in a flat $d$-dimensional Euclidean space
$Z(h)$, given by
\begin{equation}
Z(h)=\int [d\varphi]\, \exp\left(-S[\varphi]+\int d^{d}x\, h(x)\,\varphi(x)\right),
\label{e1}
\label{1}
\end{equation}
where $[d\varphi]$ is a translational invariant measure (formally 
given by  $[d\varphi]=\prod_{x\in\,R^{d}}d\varphi(x)$) and 
$S[\varphi]$ is the classical 
action associated with the scalar field. The quantity $Z(h)$ can be regarded 
as the functional integral representation for the imaginary time evolution
operator $\left<\varphi_{2}|U(t_{2},t_{1})|\varphi_{1}\right>$, with boundary 
conditions $\varphi(t_{1},\vec{x})=\varphi_{1}(\vec{x})$ and  
$\varphi(t_{2},\vec{x})=\varphi_{2}(\vec{x})$ which gives the transition 
amplitude from the initial state $|\varphi_{1}>$
to a final state  $|\varphi_{2}>$ in the presence of some scalar source 
of compact support. As usual, the generating functional 
of the connected correlation functions shall be 
given by $W(h)=\ln Z(h)$. In a 
free scalar theory, $Z(h)$ as well as $W(h)$ can be 
calculated exactly. Regarding the Lagrangian density, 
we assume that 
\begin{equation}
{\cal{L}}(\varphi,\,\partial\varphi)=\frac{1}{2}(\partial\varphi)^{2}+
\frac{1}{2}m_{0}^{2}\varphi^{2}+\frac{1}{4!}\lambda_{0}\varphi^{4},
\label{2}
\end{equation}
where $m_{0}$ is the bare mass and $\lambda_{0}$ is the bare 
coupling constant of the model. We are also assuming $m_{0}^{2}\geq 0$
and also $\lambda_{0}>0$.
The Euclidean $n$-point correlation functions, i.e., the $n$-point Schwinger functions are given by the expectation value with respect to the weight $\exp(-S(\varphi))$, defined as 
\begin{equation}
G^{(n)}(x_{1},x_{2},.,x_{n})=
\frac{1}{Z(h)}\frac{\delta^{n}\,Z(h)}{\delta h(x_{1})..\delta h(x_{n})}|_{h=0}.
\label{3}
\end{equation}
The $n$-point connected correlation functions $G^{(n)}_{c}(x_{1},x_{2},.,x_{n})$ are given by 
\begin{equation}
G^{(n)}_{c}(x_{1},x_{2},.,x_{n})=
\frac{\delta^{n}\,W(h)}{\delta h(x_{1})..\delta h(x_{n})}|_{h=0}.
\label{4}
\end{equation}
Finally, the generating functional 
of connected one-particle irreducible correlation 
functions (the effective action) is introduced by performing a Legendre transformation on $W(h)$, 
\begin{equation}
\Gamma(\varphi_{0})=-W(h)+\int\,d^{d}x\,\varphi_{0}(x)h(x).
\label{5}
\end{equation}
Let us define the proper vertices $\Gamma^{(n)}(x_{1},..,x_{n})$ as:
\begin{equation}
\Gamma^{(n)}(x_{1},..,x_{n})=\frac{\delta^{n}
\Gamma(\varphi_{0})}{\delta\varphi_{0}
(x_{1}),..\delta\varphi_{0}(x_{n})}|_{\varphi_{0}=0},
\label{6}
\end{equation}
where the normalized vacuum expectation value of the field $\varphi_{0}(x)$ 
is given by
\begin{equation}
\varphi_{0}(x)=\frac{\delta W}{\delta h(x)}.
\label{7}
\end{equation}
It is clear that in the case of a single scalar field,
for a zero normalized vacuum expectation value of the 
field $\varphi_{0}(x)$, the effective action 
may be represented as a functional power series around the value $\varphi_{0}=0$, with 
the form
\begin{equation}
\Gamma(\varphi_{0})=\sum_{n=0}^{\infty}\,\frac{1}{n!}\int\, d^{d}x_{1}..d^{d}x_{n}
\,\Gamma^{(n)}(x_{1},...,x_{n})
\varphi_{0}(x_{1})...\varphi_{0}(x_{n}).
\label{8}
\end{equation}
If the bare coupling constant vanishes, i.e., $\lambda_{0}
=0$, the generating functional of all $n$-point
Schwinger functions 
$Z(h)$ can be calculated exactly, since we have to evaluate only 
Gaussian integrals. After some manipulations we obtain that 
the Gaussian generating functional $Z_{0}(h)$ is given by
\begin{equation}
Z_{0}(h)=\exp\left(\frac{1}{2}\int\, d^{d}x\,\int\, d^{d}y\,\,h(x)\,G_{0}^{(2)}(x-y,m_{0})h(y)\right),
\label{9}
\end{equation}
where the two-point Schwinger function (the inverse kernel) satisfies
\begin{equation}
(-\Delta_{x}+m_{0}^{2})G_{0}^{(2)}(x-y,m_{0})=\delta^{d}(x-y).
\label{10}
\end{equation}
In this situation, the free Euclidean field is a gaussian 
random variable defined by its two-point 
correlation function
\begin{equation}
G_{0}^{(2)}(x-y,m_{0})=\left<x|(-\Delta+m_{0}^{2})^{-1}|y\right>,
\label{11}
\end{equation}
and the two-point Schwinger function has a well 
known Fourier representation given by
\begin{equation}
G_{0}^{(2)}(x-y,m_{0})=\frac{1}{(2\pi)^{d}}\int\, d^{d}p\frac{e^{ip(x-y)}}{(p^{2}+m_{0}^{2})}.
\label{12}
\end{equation}

In the next chapter we will show that the two-point function 
$G_{0}^{(2)}(x-y,m_{0})$ can be expressed 
in terms of the modified Bessel 
function of the third kind or Macdonald's function 
$K_{\mu}(x)$. At present, we are not interested in evaluating 
the two-point Schwinger function, but only in the analysis of 
the behavior of $G_{0}^{(2)}(x-y,m_{0})$ in a given $\epsilon$ 
-neighborhood. Let us assume that $m|x-y|<<1$; in this case,
for $d\geq 3$ we can use that 
$G_{0}^{(2)}(x-y,m_{0}^{2})\approx\,G_{0}^{(2)}(x-y,m_{0}^{2}=0)=|x-y|^{-(d-2)}$. For $d=2$, the mass parameter can 
not be eliminated from the denominator and we have the following short 
distance behavior: $G_{0}^{(2)}(x-y,m_{0}^{2})\propto\, \ln(m|x-y|)$. 
It is well known that a massless two-dimensional scalar
field theory is not consistent, once the model has severe
infrared divergences. There are different proposals to circumvent 
this problem; we only mention some of them. 
For instance, one may violate the 
positivity of the state vector space; another attempt is to restrict the test functions of the theory, and finally one can introduce  a cut-off in the 
definition of the positive and negative Wightman functions. It is clear that
such cut-off procedure is equivalent to introducing a box to regulate the theory in the infrared.  
Later, we will discusss other strategies to solve the problem of the 
infrared divergences in scalar theories at finite temperature.

Coming back to the generating functional of all Schwinger functions,
for $\lambda_{0}\neq\,0$ it is not possible 
to find a closed exact expression for
the partition function, and a perturbative expansion is mandatory.
Let us then assume the weak-coupling perturbative expansion of the 
theory. It is important 
to point out that the partition function
can be defined in arbitrary geometries, and classical 
boundary conditions must be implemented 
in the two-point Schwinger function, restricting the space of 
functions that appear in the functional integrals.
If we want to include thermal effects, and assuming thermal equilibrium, from  
the Feynman, Matheus and Salam formula we 
have:
\begin{equation}
\left<\varphi_{b}|e^{-iH(t_{f}-t_{i})}|\varphi_{a}\right>=
\int_{\varphi(t_{i})=\varphi_{a}}^{\varphi(t_{f})=\varphi_{b}}
\exp\left(i\int_{t_{i}}^{t_{f}}\,dt\,\int\,d^{d-1}x\,{\cal L}(\varphi,\,\partial\varphi)\right),
\label{13}
\end{equation}
where we have to assume that $t_{f}-t_{i}=-i\beta$ and also set $\varphi_{a}=\varphi_{b}$, and the 
sum over all $\varphi_{a}$ must be performed in order to produce the 
trace. The partition function $Tr\left[e^{-\beta\,H}\right]$ is given by
\begin{equation}
Tr\left[e^{-\beta\,H}\right]=\int_{periodic}\,[d\varphi]
\exp\left(i\int_{t{i}}^{t_{i}-i\beta}dt\,\int d^{d-1}x\,{\cal L}
(\varphi,\,\partial\varphi)
\right),
\label{14}
\end{equation}
where the integration over the fields 
satisfying $\varphi(t_{i}-i\beta,{\vec {x}})=\varphi(t,{\vec{x}})$.
Since the time integration must range from some value $t_{i}$ to $t_{i}-i\beta$, let $t_{i}=0$ and 
set the contour along the negative imaginary axis from $0$ to $-i\beta$. Thus, 
$t=-i\tau$, where $0\leq\,\tau\,\leq \beta$, and we have
\begin{equation}
Z(h)|_{h=0}=\int_{periodic}\,[d\varphi]
\exp\left(\int_{0}^{\beta}d\tau\,\int d^{d-1}x\,{\cal L}(\varphi,\partial\varphi)\right).
\label{15}
\end{equation}
To generate the $n$-point Schwinger functions we 
need to couples the field with an external source.
We will assume that the system is confined between  two paralel hyperplates, 
(which we call the Casimir configuration), localized at $z=0$ and $z=L$, and we are using 
cartesian coordinates $x^{\mu}=({\vec{r}},z)$, where $\vec{r}$ is a 
$(d-1)$ dimensional vector 
perpendicular to the $\vec{z}$ vector. Note that 
since we assume thermal equilibrium with a reservoir,
we have periodicity in the first coordinate and
$0\leq r_{1}\leq \beta$. See for example Ref. \cite{bernard}, or 
for a complete review of quantum field theory 
at thermal equilibrium, see for example Ref. \cite{kaplivro}.
The choice of Dirichlet-Dirichlet boundary conditions means that the 
scalar field satisfies 
\begin{equation}
\varphi({\vec{r}},z)|_{z=0}=\varphi({\vec{r}},z)|_{z=L},
\label{16}
\end{equation}
and
Neumann-Neumann boundary conditions means that 
\begin{equation}
\frac{\partial}{\partial z}\varphi({\vec{r}},z)|_{z=0}=
\frac{\partial}{\partial z}\varphi({\vec{r}},z)|_{z=L}.
\label{16a}
\end{equation}

In the next section we will discuss the perturbative 
renormalization at the one-loop level
of the field theory in the presence of rigid boundaries.
The great interest of this matter is: 
when systems contain macroscopic structures, how it 
is possible to implement the renormalization program?
We will examinate how does the weak-coupling perturbative expansion 
and the renormalization program can be implemented.
In order to identify the singular part of the one-loop two-point  
Schwinger function, we use a combination of dimensional and zeta-function 
analytic regularization procedures. We also present the detailed form of 
the surface divergences. Note that due to our choice (two-parallel 
hyperplates), the region outside the boundaries is the 
union of two-simple connected domains. The renormalization 
of the field theory in such exterior regions  must be carried out 
along the same lines as for the interior region. For simplicity we are 
considering only the interior region.

\section{The regularized one-loop two and four-point Schwinger functions}

The aim of this section is to reshape a well known result, adding 
finite temperature effects to the problem.
In order to implement the renormalization program in 
a scalar field theory where we assume 
Dirichlet-Dirichlet or Neumann-Neumann boundary conditions on  
rigid surfaces one has to introduce surface counterterms. 
To write the full renormalized action for the theory with rigid boundaries 
we need two regulators: the first one is the usual $\epsilon$ that is introduced in 
the dimensional regularization procedure and the second one 
which we call $\eta$, represents 
the distance to a boundary.  
Accordingly we will show that the full renormalized action must be given by:
\begin{eqnarray}
S(\varphi)& &= 
\int_{0}^{L}dz \int d^{d-1}r
\left(\frac{A(\epsilon)}{2}(\partial_
{\mu}\varphi)^{2}+\frac{B(\epsilon)}{2}\varphi^2+
\frac{C(\epsilon)}{4!}\varphi^{4}\right)\nonumber\\
& &+
\int d^{d-1}r\left(c_{1}(\eta)\varphi^{2}(\vec{r},0)+c_{2}(\eta)
\varphi^{2}(\vec{r},L)\right)\nonumber\\
& &+
\int d^{d-1}r\left(c_{3}(\eta)\varphi^{4}(\vec{r},0)+c_{4}(\eta)
\varphi^{4}(\vec{r},L)\right),
\label{sur}
\end{eqnarray}
where $A(\epsilon)$, $B(\epsilon)$ and $C(\epsilon)$ are 
the usual coefficients for the bulk counterterms and 
the coefficients $c_{i}(\eta)$, $i=1,..4$, which 
depend on the boundary conditions for the field, 
are the coefficients for the surface 
counterterms. As usual, all of these coefficients   
must be calculated order by order 
in  perturbation theory. Note that we are interested in
systems that are invariant under translation
along directions parallel to the plates, which
implies that the full momentum is not conserved. For such conditions, 
a more convenient 
representation for the n-point Schwinger functions
to implement the perturbative renormalization
is a mixed $(\vec{p}, z)$
representation. Careless one-loop perturbation theory leads to 
ultraviolet counterterms that depend on the distance 
between the plates or also to 
the absence of surface counterterms \cite{erro1} \cite{erro2}.

In a straightforward way,
in the Matsubara formalism all the Feynman rules are the same 
as in the zero temperature case, except that the momentum-space
integrals over the zeroth component is replaced by a sum over
discrete frequencies. For the case of bosons fields we have to 
perform the replacement
\begin{equation}
\int \frac{d^{d}p}{(2 \pi)^d}f(p)\rightarrow \frac{1}{\beta} \sum_{n} \int \frac{d^{d-1}p}{(2 \pi)^{d-1}}f(\frac{2n\pi}{\beta}, \vec{p}),
\label{sub}
\end{equation}
where we are using the following notation: $(\int d^{d-1}r=\int_{0}^{\beta}
dr_{1}\int d^{d-2}r)$.

We begin the study of the interacting theory by building the
one-loop correction $\left(G_{1}^{(2)}(\lambda_{0},x,x')\right)$ to the bare two-point
Schwinger function $G_{0}^{(2)}(x,x')$, for both the $DD$ and $NN$ 
boundary conditions. Using the Feynman rules we have that 
$G_{1}^{(2)}(\lambda_{0},\vec{r}_{1},z_{1},\vec{r}_{2},z_{2})$ can be 
written as
\begin{equation}
G_{1}^{(2)}(\lambda_{0},\vec{r}_{1},z_{1},\vec{r}_{2},z_{2})
=\frac{\lambda_{0}}{2}\int{}d^{d-1}r\int_{0}^{L}dz
\,\,G_{0}^{(2)}(\vec{r}_{1}-\vec{r},z_{1},z)\,
G_{0}^{(2)}(\vec{0},z)\,
G_{0}^{(2)}(\vec{r}-\vec{r}_{2},z,z_{2}).
\label{MF}
\end{equation}
Even though the functions
$G_{0}^{(2)}(\vec{r}_{1}- \vec{r}_{2},z_{1},z_{2})$ and
$G_{0}^{(2)}(\vec{r}_{2}-\vec{r}_{3},z_{2},z_{3})$ are singular at
coincident points ($\vec{r}_{1}=\vec{r}_{2}$, $z_{1}=z_{2}$) and
($\vec{r}_{2}=\vec{r}_{3}$, $z_{2}=z_{3}$), the singularities are
integrable for points outside the plates.
It is worth mentioning that the most simple way to take into account the boundary is to implement the boundary conditions through the 
explicit form of the free two-point Schwinger function
$G_{0}^{(2)}(x-y,m_{0})$.
A straightforward substitution yields the order $\lambda_{0}$
correction to the bare two-point Schwinger function in the one-loop
approximation for the case of Dirichlet-Dirichlet boundary conditions.
Using the Feynman rules,
$G_{2}^{(4)}(\lambda_{0},x_{1},x_{2},x_{3},x_{4})$, i.e.,  
the $O(\lambda_{0}^{2})$
correction to the bare one-loop four-point Schwinger function, 
is given by
\begin{eqnarray}
G_{2}^{(4)}(\lambda_{0},\vec{r}_{1},z_{1},\vec{r}_{2},z_{2},
\vec{r}_{3},z_{3},\vec{r}_{4},z_{4})
&=& \frac{\lambda_{0}^2}{2}\int{}d^{d-1}r\int{}d^{d-1}r^{\prime}
\int_{0}^{L}dz \int_{0}^{L}dz^{\prime}
\;G_{0}^{(2)}(\vec{r}_{1}-\vec{r},z_{1},z)
\nonumber \\
&&G_{0}^{(2)}(\vec{r}_{2}-\vec{r},z_{2},z)
\left(G_{0}^{(2)}(\vec{r}-\vec{r^{\prime}},z,z^{\prime})\right)^2\nonumber\\
&& G_{0}^{(2)}(\vec{r^{\prime}}-\vec{r}_{3},z^{\prime},z_{3})
G_{0}^{(2)}(\vec{r^{\prime}}-\vec{r}_{4},z^{\prime},z_{4}).
\label{nova}
\end{eqnarray}
Note that we supress the $m_{0}$ term in each expression.
Again, all $G_{0}$'s are singular at coincident points, but the
singularities are integrable for points outside the 
plates, except for
$G_{0}^{(2)}(\vec{r}-\vec{r^{\prime}},z,z^{\prime})$.
Having in mind the above discussion,
in this section we will study the 
following expressions:
\begin{equation}
\frac{\lambda_{0}}{2}\int{}d^{d-1}r\int_{0}^{L}dz
\,\,
\left(G_{0}^{(2)}(\vec{0},z)\right),
\label{17}
\end{equation}
and
\begin{equation}
\frac{\lambda_{0}^2}{2}\int{}d^{d-1}r\int{}d^{d-1}r^{\prime}
\int_{0}^{L}dz \int_{0}^{L}dz^{\prime}
\left(G_{0}^{(2)}(\vec{r}-\vec{r^{\prime}},z,z^{\prime})\right)^2.
\label{18}
\end{equation}
Let us first study $\frac{1}{2}\,G_{0}^{(2)}(\vec{0},z)
\equiv I(z,m_{0},L,\beta,d)$, and define
the following quantities: $\frac{1}{b}=\frac{2}{\beta}$, $L=a$ 
and finally the dimensionless coupling constant $g=\mu^{4-d}\lambda_{0}$.
Therefore, the argument in the 
integral defined in Eq.(\ref{17}), 
$I(z,m_{0},a,b,d)$ can be written as
\begin{equation}
I(z,m_{0},a,b,d)=\frac{g}{2(2\pi )^{d-2}ab}
\sum_{n=-\infty}^{\infty}\sum_{n^{\prime
}=1}^{\infty}\sin ^{2}( \frac{n^{\prime}\pi z}{a})
\int d^{d-2}p\frac{1}{\left( \vec{p}^{\,2}+
(\frac{n^{\prime }\pi }{a})^{2}+(\frac{n\pi }{b})^{2}+
+m_{0}^{2}\right) }.  
\label{19}
\end{equation}
There are two points that we would like to stress. First 
to perform analytic regularizations we 
have to introduce a parameter $\mu$ with dimension of mass in order to 
have dimensionless quantities raised to a complex power.
Second, the generalization for the case of Neumann 
boundary conditions is 
straightforward, although in this case infrared 
divergences associated with the 
$n=0$ mode will appear in the case of massless scalar 
field. To circumvent this situation, 
we must have a finite Euclidean volume to regularize 
the model in the infrared, or trying to implement a resummation to generate 
a thermal mass. We will return to this point latter.

Using trigonometric identities, it is convenient to write the 
amputated one-loop two-point Schwinger 
in two parts. The first comprises the contributions that do 
not depend on the distance 
to the boundary, and the second   
the contributions that do depend on this distance.
Therefore, the 
quantity $I(z,m_{0},a,b,d)$ can be split in 
two parts $T_{1}(m_{0},a,b,d)$ and $T_{2}(z,m_{0},a,b,d)$, i.e.: 
\begin{equation}
I(z,m_{0},a,b,d)=T_{1}(m_{0},a,b,d)+T_{2}(z,m_{0},a,b,d).
\label{20}
\end{equation}
The first 
quantity $T_{1}(m_{0},a,b,d)$,
independent on the distance to the boundaries can be expressed
in the following way: 
\begin{equation}
T_{1}(m_{0},a,b,d)=I_{0}(m_{0},a,b,d)+I_{1}(m_{0},a,b,d)+I_{2}(m_{0},a,b,d),
\label{21}
\end{equation}
where each term is given respectivelly by:
\begin{equation}
I_{0}(m_{0},a,b,d)=-\frac{g}{16(2\pi )^{d-2}ab}\int d^{d-2}p\frac{
1}{(\vec{p}^{\,2}+m_{0}^{2})},  
\label{22}
\end{equation}
\begin{equation}
I_{1}(m_{0},a,b,d)=\frac{g}{8\left( 2\pi \right) ^{d-2}ab}
\sum_{n=1}^{\infty }\int d^{d-2}p\frac{1}{\left(\vec{p}^{\,2}+m_{0}^{2}+
(\frac{n\pi }{a})^{2}\right)},
\label{23}
\end{equation}
and finally
\begin{equation}
I_{2}(m_{0},a,b,d)=\frac{g}{4(2\pi )^{d-2}ab}\sum_{n,n^{\prime }=1}^{\infty
}\int d^{d-2}p\frac{1}{\left( \vec{p}^{\,2}+(\frac{n\pi }{a})^{2}+(\frac{
n^{\prime }\pi }{b})^{2}+m_{0}^{2}\right)}.  
\label{24}
\end{equation}
The contribution that depends on the distance to the boundaries given by
$T_{2}(z,m_{0},a,b,d)$, can be split in the following way:
\begin{equation}
T_{2}(z,m_{0},a,b,d)=I_{3}(z,m_{0},b,d)+I_{4}(z,m_{0},a,b,d)+
I_{5}(z,m_{0},b,d)+
I_{6}(z,m_{0},a,b,d).
\label{25}
\end{equation}
Each term contributing to $T_{2}(z,m_{0},a,b,d)$ is given, 
respectivelly by:
\begin{equation}
I_{3}(z,m_{0},b,d)=\frac{g}{2b}h(d)\int_{m_{0}}^{\infty}\,dv
(v^{2}-m_{0}^{2})^{\frac{d-4}{2}}\exp(-2vz),
\label{26}
\end{equation}
\begin{equation}
I_{4}(z,m_{0},a,b,d)=\frac{g}{2b}h(d)\int_{m_{0}}^{\infty}\,dv\,
(v^{2}-m_{0}^{2})^{\frac{d-4}{2}}(\coth av-1)\cosh 2vz,
\label{27}
\end{equation}
\begin{equation}
I_{5}(z,m_{0},b,d)=\frac{g}{b}h(d)\sum_{n=1}^{\infty}
\int_{m_{0}}^{\infty}\,dv\,
\left(v^{2}-m_{0}^{2}-(\frac{n\pi}{b})^{2}\right)^{\frac{d-4}{2}}
\exp(-2vz),
\label{28}
\end{equation}
and finally 
\begin{equation}
I_{6}(z,m_{0},a,b,d)=\frac{g}{b}h(d)\sum_{n=1}^{\infty}
\int_{\alpha}^{\infty}\,dv\,
\left(v^{2}-m_{0}^{2}-(\frac{n\pi}{b})^{2}\right)^{\frac{d-4}{2}}
(\coth av-1)\cosh 2vz.
\label{29}
\end{equation}
In the above expression the quantity $\alpha$ is given by
\begin{equation}
\alpha=\left(m_{0}^{2}+(\frac{n\pi}{b})^{2}\right)^{\frac{1}{2}},
\label{30}
\end{equation}
and $h(d)$, that appears in Eqs.(\ref{26}), (\ref{27}), (\ref{28}) and 
(\ref{29}) is an entire function given by
\begin{equation}
h(d)=\frac{1}{4(4\pi)^{\frac{d-2}{2}}}\frac{1}{\Gamma(\frac{d-2}{2})}.
\label{31}
\end{equation}
Let us investigate each contribution 
in detail. Using dimensional regularization we obtain for 
$I_{0}(m_{0},d)$ the following expression:
\begin{equation}
I_{0}(m_{0},a,b,d)=-\frac{g}{16\,ab(2\sqrt{\pi})^{d-2}}\Gamma(2-\frac{d}{2})
(m_{0}^{2})^{\frac{d}{2}-2}.
\label{32}
\end{equation}
An analytic expression for the Gamma function $\Gamma(z)$, defined
in the whole 
complex plane, can be found and in the neighborhood of a pole $z=-n$, $(n=0,1,2..)$ 
the Gamma function has the representation
\begin{equation}
\Gamma(z)=\frac{(-1)^{n}}{n!}\frac{1}{(z+n)}+\Omega(z+n),
\label{33}
\end{equation}
with regular part $\Omega(z+n)$. Using that $4-d=\epsilon$ an the duplication formula 
for the Gamma function $\Gamma(z)$ we have 
\begin{equation}
I_{0}(m_{0},a,b,d)|_{d=4}=-\frac{g}{16\pi\,ab}\frac{1}{m^{\epsilon}_{0}}
\left(\frac{1}{\epsilon}+\Omega(\epsilon)\right).
\label{34}
\end{equation}
Here one may adopt different renormalization schemes. 
We can choose the minimal subtraction (MS) scheme, 
in which we eliminate only the pole term $\frac{1}{\epsilon}$ 
in the dimensionaly regularized expression for the Schwinger functions. 
Another choice is the modified minimal subtraction (MS) scheme,
where we 
eliminate not only the pole term $\frac{1}{\epsilon}$ but 
also the regular part around the pole. Note that in the minimal subtraction scheme 
the counterterms acquire the simplest expression, while the renormalized 
Schwinger functions have more complicated expressions.
Let us analyse the second expression, given by $I_{1}(m_{0},a,b,d)$. Using dimensional 
regularization it is possible to show that
\begin{equation}
I_{1}(m_{0},a,b,d)=\frac{g}{8(2\sqrt\pi)^{d-2}ab}\Gamma (2-
\frac{d}{2})\sum_{n=1}^{\infty }\frac{1}{{\left( m_{0}^{2}+(\frac{
n\pi }{a})^{2}\right) ^{2-\frac{d}{2}}}}.
\label{35}
\end{equation}
We note that to extract a finite result from $I_{1}(m_{0},a,b,d)$ we still 
have to use the analytic extension of the Epstein-Hurwitz zeta function. A direct 
calculation gives
\begin{eqnarray}
I_{1}(m_{0},a,b,d)&=&-\frac{g}{8ab}m_{0}^{d-4}
\frac{\sqrt{\pi}}{(2{\sqrt{\pi}})^{d-1}}\Gamma(2-\frac{d}{2})+
\nonumber\\
&&
\frac{g\,m_{0}^{d-3}}{8b}\frac{1}{(2\pi)^{d-1}}\left
(\Gamma\left(\frac{3-d}{2}\right)+
4\sum_{n=1}^{\infty}\,(am_{0}n)^{\frac{3-d}{2}}
K_{\frac{3-d}{2}}(2m_{0}na)
\right).
\label{36}
\end{eqnarray} 
The first term in the above equation is a polar part and the second one is finite. 
Assuming the minimal subtraction scheme, $I_{1}(m_{0},a,b,d)$ becomes finite.
The next term that we have to analyse is $I_{2}(m_{0},a,b,d)$ defined by:
\begin{equation}
I_{2}(m_{0},a,b,d)=\frac{g}{4ab}\frac{1}
{(2\pi )^{d-2}}
\sum_{n,n^{\prime }=1}^{\infty }\int d^{d-2}p\frac{1
}{\left( \vec{p}^{\,2}+(\frac{n\pi }{a})^{2}+(\frac{n^{\prime }\pi }{b}
)^{2}+m_{0}^{2}\right) }.  
\label{37}
\end{equation}
The contribution given by the above equation
is a part of the amputated one-loop two-point Schwinger function
that does not depend on the distance to the boundaries, 
and in the renormalization 
procedure it will require only a usual bulk counterterm. The form of 
the counterterm is given by the principal part of the Laurent expansion 
of Eq.(\ref{37}) around some $d$, which must be given by the analytic 
extension of the Epstein zeta function in the complex $d$ plane. 
The structure of the divergences of the Epstein zeta function is well known 
in the literature \cite{ambb} \cite{ford} \cite{ori} \cite{kirst}. 
Since the polar structure of the above equation 
can be found in the literature, we will focus only on the 
position-dependent divergent part given by $T_{2}(z,m_{0},a,b,d)$. 
We are now in position to discuss the behavior of 
$I_{3}(z,m_{0},b,d),I_{4}(z,m_{0},a,b,d),I_{5}(z,m_{0},b,d)$ and finally $I_{6}(z,m_{0},a,b,d)$.

Let us first analyse $I_{3}(z,m_{0},b,d)$.
Using the following integral 
representation of the 
modified Bessel functions of third kind, or Macdonald's 
functions $K_{\nu}(x)$ \cite{grad},
\begin{equation}
\int_{u}^{\infty}(x^{2}-u^{2})^{\nu-1}e^{-\mu x} \,dx=
\frac{1}{\sqrt{\pi}}(\frac{2u}{\mu})^{\nu-\frac{1}{2}}\Gamma(\nu)
K_{\nu-\frac{1}{2}}(u\mu),
\label{38}
\end{equation}
which is valid for $u>0$, $Re\,(\mu)>0$ and $Re\,(\nu)>0$, 
we see that $I_{3}(z,m_{0},a,b,d)$ 
can be written in terms of these functions. A simple substitution gives
\begin{equation}
I_{3}(z,m_{0},a,b,d)=\frac{2}{b}\frac{h(d)}{(2\sqrt{\pi})^{d-1}}
(\frac{m_{0}}{z})^{\frac{d-3}{2}}
K_{\frac{d-3}{2}}(2m_{0} z).
\label{39}
\end{equation}
Using a asymptoptic formula for the Bessel function,  
$I_{3}(z,m_{0},a,b,d)$ is given by
\begin{equation}
I_{3}(z,m_{0},a,b,d)=\frac{2}{b}\frac{h(d)}{(2\sqrt{\pi})^{d-1}}
\frac{\Gamma(\frac{d-3}{2})}{z^{d-3}}.
\label{40}
\end{equation}
We can see that we have a divergent behavior as $z\rightarrow 0$, which 
demands 
a surface counterterm. Let us show that the other terms also contain 
surface divergences, and study $I_{4}(z,m_{0},a,b,d)$.
To advance in the calculations, we have to extend the binomial series for 
both positive or negative integral exponents, written in the 
form
\begin{equation}
(1+x)^{k}=\sum_{n=0}^{\infty}C_{n}^{k}\,x^{n}.
\label{41}
\end{equation} 
First, it is possible 
to show that the binomial expansion holds for any real exponent
$\alpha$, $|x|<1$ and $\alpha\, \epsilon \,\,R $, i.e.,
\begin{equation}
(1+x)^{\alpha}=\sum_{n=0}^{\infty}C_{\alpha}^{n}\,x^{n},
\label{42}
\end{equation}
where $C_{\alpha}^{n}$ are the generalization of the binomial coefficients. Since we are 
using dimensional regularization, it is possible to extend the binomial expansion when 
both the exponent $\alpha$ as well 
the variable $x$ assume complex values. For this 
purpose we use the following theorem:

For any complex exponent $\alpha$ and any complex $z$ in $|z|<1$, the 
binomial series
\begin{equation} 
\sum_{n=0}^{\infty}C_{\alpha}^{n}z^{n}=1+C_{\alpha}^{1}z+..+
C_{\alpha}^{n}z^{n}+..
\label{43}
\end{equation}
converges and has for 
sum the principal value of the power $(1+z)^{\alpha}$, 
where the principal value of the power $b^{a}$ is 
given by the number uniquely defined by the formula 
$b^{a}=\exp(a\ln b)$, where $a$ and $b$ denotes any 
complex numbers, with $b\neq 0$ as the only condition, 
and $\ln b$ is given its principal value. Going back to  
$I_{4}(z,m_{0},a,b,d)$, using the 
generalization of the binomial theorem, let us define
$C^{(1)}(d,k)=\frac{1}{2}h(d)(-1)^{k}C_{\frac{d-4}{2}}^{k}$ to obtain
\begin{equation}
I_{4}(z,m_{0},a,b,d)=\frac{g}{a^{d-3}b}\sum_{k=0}^{\infty}C^{(1)}(d,k)(m_{0}a)^{2k}
\int_{m_{0}a}^{\infty}\,u^{d-4-2k}(\coth u-1)\cosh(\frac{2uz}{a}).
\label{44}
\end{equation}
Let us use the following integral representation 
of the Gamma function,
\begin{equation}
\int_{0}^{\infty} dt \,t^{\mu-1}e^{-\nu
t}=\frac1{\nu^{\mu}}\Gamma(\mu), \,\,\,\,Re(\mu)>0,\,\,\,Re(\nu)>0,
\label{45}
\end{equation}
and also the following integral representation
of the product of the Gamma function times the Hurwitz zeta function 
\begin{equation}
\int_{0}^{\infty} dt \,t^{\mu-1}e^{-\alpha t}(\coth
t-1)=2^{1-\mu}\Gamma(\mu) 
\zeta(\mu,\frac{\alpha}{2}+1)\,\,\,\,Re(\alpha)>0,\,
\,\,Re(\mu)>1,  
\label{46}
\end{equation}
where $\zeta(s,u)$ is the Hurwitz zeta function defined by \cite{grad} 
\begin{equation}
\zeta(s,u)=\sum_{n=0}^{\infty}\frac{1}{(n+u)^{s}},\,\,\,\,Re(s)>1
\,\,\,\,\, u \neq 0,-1,-2...  
\label{47}
\end{equation}
It is not difficult to show that $I_{4}(z,m_{0},a,b,d)$ contains 
surface divergences at $z=0$ and also $z=a$. For more details, see for 
example Ref. \cite{polarization}. The other expression that we have 
to study is
$I_{5}(z,m_{0},a,b,d)$. Using an integral representation of the 
Bessel function of third kind we have:
\begin{equation}
I_{5}(z,m_{0},a,b,d)=\frac{1}{b}\frac{1}{(2\sqrt{\pi})^{d-1}}
\sum_{n=1}^{\infty}
(\frac{\alpha}{z})^{\frac{d-3}{2}}
K_{\frac{d-3}{2}}(2\alpha z).
\label{48}
\end{equation}
Using an asymptotic representation of the Bessel 
function it is posssible to 
present also the singular behavior near $z=0$.
Let us finally investigate $I_{6}(z,m_{0},a,b,d)$. 
A simple calculation for the massless case gives
\begin{equation}
I_{6}(z,m_{0},a,b,d)|_{m=0}=
\frac{1}{a^{d-3}b}\sum_{k=0}^{\infty}C^{(2)}(d,k)(\frac{a}{b})^{2k}
\sum_{n=1}^{\infty}
n^{2k}\int_{\frac{n\pi a}{b}}^{\infty}du\,
u^{d-4-2k}(\coth u-1)\cosh (\frac{2uz}{a}),
\label{49}
\end{equation}
where $C^{(2)}(d,k)=h(d)(-1)^{k}C_{\frac{d-4}{2}}^{k}\pi^{2k}$ is an 
entire function in the complex $d$ plane.
The integral that appear in Eq.(\ref{49})
cannot be evaluated explicity in terms of well 
known functions. Nevertheless it is possible to write Eq.(\ref{49}) 
in a convenient way where the structure of the divergences near the plate 
when $y\rightarrow b$ appear. Clearly for details see Ref. \cite{polarization}.
In the next section we will investigate the singularities of the four-point Schwinger function.
 
\section{The four-point Schwinger function in the one-loop approximation}

We now turn our attention to the four-point Schwinger function in the 
one-loop approximation. For simplicity we shall study only the 
zero temperature case. In this section we are 
following the discussion developed in Ref. \cite{caicedo}.
Introducing new variables as
$u_\pm{}\equiv{}z\pm{}z'$, and also $(\vec{\rho}=\vec{r}-\vec{r'})$,
the  zero-temperature two-point Schwinger function in the tree-level 
$G_{0}^{(2)}(\vec{\rho},z,z')$ can be split into
\begin{equation}
G_{0}^{(2)}(\vec{\rho},z,z')=G_{+}^{(2)}(\vec{\rho},u_+)+
G_{-}^{(2)}(\vec{\rho},u_-),
\label{split-of-G}
\end{equation}
where we are defining $A_{n}(a,m_{0},d,\vec{\rho})$ by
\begin{equation}
A_{n}(a,m_{0},d,\vec{\rho})=\frac{1}{(2\pi)^
{d-1}}\int{}d^{d-1}p\frac{e^{i\vec{p}.\vec{\rho}}}
{(\vec{p}^{\,2}+(\frac{n\pi}{a})^{2}+m_{0}^{2})},
\label{v}
\end{equation}
and so $G_{\pm}^{(2)}(\vec{\rho},u_\pm)$ can be expressed as
\begin{equation}
G_{\pm}^{(2)}(\vec{\rho},u_\pm)=
\mp\frac{1}{a}\sum_{n=1}^{\infty}
\cos(\frac{n\pi u_\pm}{a})A_{n}(a,m_{0},d,\vec{\rho}).
\label{D-s}
\end{equation}
Before proceeding, let us 
present a explicit formula for the  
free two-point Schwinger function $G_{\pm}^{(2)}(\rho,u_\pm)$
in terms of Bessel functions. 
Let us define an analytic function $f(d)$ by 
\begin{equation}
f(d)=\frac{1}{\sqrt{\pi}(2\pi)^{\frac{d-1}{2}}}
\frac{\Gamma(\frac{d-2}{2})}{\Gamma(\frac{d-3}{2})}. 
\end{equation}
Strictly speaking, it is possible to show that we can write 
$G_{\pm}^{(2)}(\rho,u_\pm)$ in terms of the Bessel function of 
third kind. To this end, we use the standard formula
\begin{equation}
\frac{1}{(2\pi)^d}
\int d^dr F(r)e^{i\vec{k}.\vec{r}}=
\frac{1}{\sqrt{\pi}(2\pi)^{\frac{d}{2}}}
\frac{\Gamma(\frac{d-1}{2})}{\Gamma(\frac{d-2}{2})}
\int_0^{\infty}F(r)r^{\frac{d}{2}}J_{\frac{d-3}{2}}(kr)dr,
\end{equation}
which leads us to:
\begin{equation}
G_{\pm}^{(2)}(\rho,u_\pm)=
\mp\frac{f(d)}{\rho^{\frac{d-3}{2}}a}
\sum_{n=1}^\infty{}\cos(\frac{n\pi{}u_\pm}{a})
\int_0^\infty dp
\frac{{p}^{\frac{d-1}{2}}}
{(p^2+(\frac{n\pi}{L})^2+m_{0}^2)}J_{\frac{d-3}{2}}(p\rho),
\label{oque}
\end{equation}
where $J_{\nu}(x)$ is the Bessel function of the first kind of order
$\nu$. The integral in Eq.(\ref{oque}) can be calculated by 
using the result \cite{grad}
\begin{equation}
\int_0^\infty dx\frac{x^{\nu+1}J_{\nu}(ax)}{(x^2+b^2)}=b^{\nu}K_{\nu}(ab),
\end{equation} 
implying that it is possible to write $G_{\pm}^{(2)}(\rho,u_\pm)$ as 
\begin{equation}
G_{\pm}^{(2)}(\rho ,u_\pm)=\mp\frac{f(d)}{\rho^{\frac{d-3}{2}}a}
\sum_{n=1}^\infty{}\cos(\frac{n\pi{}u_\pm}{a})
\left((\frac{n\pi}{a})^2+m_{0}^2\right)^\frac{d-3}{4}
K_{\frac{d-3}{2}}
\left(\rho\sqrt{m_{0}^2+(\frac{n\pi}{a})^2}\right).
\end{equation}
Using Eq.(\ref{split-of-G}) and the  
above formula, the explicit expression for the two-point Schwinger 
function in a generic $d$-dimensional Euclidean space 
confined between two flat paralel 
hyperplanes, where we assume Dirichlet-Dirichlet boundary conditions is given. 
It is difficult 
to use the above expressions for $G_{\pm}^{(2)}(\rho ,u_\pm)$ to 
investigate the analytic structure of the four-point function 
for both the bulk and near the boundaries. Nevertheless,
it is clear that the divergences of 
the four-point 
function in the one-loop approximation appear at coincident points and therefore 
the singular behavior is encoded in the polar part of 
$M(\lambda_{0},a,m,d)$ given by
\begin{equation}
M(\lambda_{0},a,m_{0},d)=
g^{2}\int d^{d-1}r \int d^{d-1}r^{\prime}
\int_{0}^{a}dz \int_{0}^{a}dz^{\prime} F(\vec{r},\vec{r}',z,z')
\left(G_{0}^{(2)}(\vec{r}-\vec{{r}^{\prime}},z,z^{\prime})\right)^2.
\label{amp}
\end{equation}
It is easy to show that $G^{(4)}_{2}(\lambda_{0},a,m_{0},d)_{amp}$ is given by
\begin{eqnarray}
&&G^{(4)}_{2}(\lambda_{0},a,m_{0},d)_{amp}=\nonumber\\
&&\frac{g^{2}}{2(2\pi)^{2d-2}}\int d^{d-1}r \int d^{d-1}r^{\prime} \int d^{d-1}k
\int d^{d-1}q \sum_{n=1}^{\infty}\,
\frac{e^{i\vec{\rho}.(\vec{q}-\vec{k})}}
{(\vec{q}^{2}+(\frac{n\pi}{a})^{2}+m_{0}^{2})(\vec{k}^{2}+
(\frac{n\pi}{a})^{2}+m_{0}^{2})},
\label{prova}
\end{eqnarray}
where $F(\vec{r},\vec{r}',z,z')$ is a regular function. As with the 
one-loop two-point function, it is 
not difficult to realize that the above equation has two kinds of singularities, 
those coming from the bulk and those arising from the behavior near the 
surface. As before, the behavior in the bulk 
is similar to the thermal field 
theory case and consequently we will discuss only the 
singularities arising from 
the boundaries. This 
can be done studying the polar part of 
${\tilde{M}}(\lambda_{0},a,m_{0},d)$ given by  
\begin{equation}
{\tilde{M}}(\lambda,a,m_{0},d)=
\frac{g^{2}}{2}
\int_{0}^{a}dz \int_{0}^{a}dz^{\prime}{\cal{F}}(z,z')
\left(G_{0}^{(2)}(\vec{0},z,z^{\prime})\right)^2,
\label{amp2}
\end{equation}
where ${\cal{F}}(z,z')$ is a regular function. Now,  we recall 
that the form of $G_{\pm}^{(2)}(\rho,u_\pm)|_{\rho=0}$ is given by,
\begin{equation}
G_\pm^{(2)}(\rho,u_\pm)|_{\rho=0}=
\mp\frac{1}{(2\pi)^{d-1}a}\sum_{n=1}^{\infty}
\cos({\frac{n\pi{u_\pm}}{a}})
\int {d}^{d-1}p
\frac{1}{\left(\vec{p}^2+m_{0}^2+(\frac{n\pi}{a})^2\right)},
\end{equation}
where it is not difficult to show that 
\begin{equation}
G_\pm^{(2)}(\rho,u_\pm)|_{\rho=0}=
\mp\left(-\frac{1}{2a}A_{0}(\rho,L,m_{0})|_{\rho=0}
+f_{2}(a,m_{0},d,\frac{u_{\pm}}{2})\right).
\end{equation}
In the above definition we are making use of the auxiliary function $f_{2}(a,d,m_{0},z)$ given by 
\begin{equation}
f_{2}(a,m_{0},d,z)=
\frac{1}{2(2\pi)^{d-1}}\int{d}^{d-1}p
\frac{1}{\sqrt{\vec{p}^2+m_{0}^2}}
\frac{\cosh((a-2z)\sqrt{\vec{p}^2+
m_{0}^2})}{\sinh(a\sqrt{\vec{p}^2+m_{0}^2})}.
\label{f2}
\end{equation}
Note that the amputated one-loop two-point 
Schwinger function can be decomposed in a translational 
invariant part and a translational 
invariance breaking part, given exactly by $f_{2}(a,m_{0},d,z)$.
When we sum to find the free
propagator, we end up with the following expression
\begin{equation}
G_{0}^{(2)}(\rho,z,z')|_{\rho=0}=f_{2}(a,m_{0},d,\frac{u_{-}}{2})-
f_{2}(a,m_{0},d,\frac{u_{+}}{2}).
\end{equation}
For the sake of simplicity, we will discuss only the massless case once 
the singularities of the massive case have the same structure as in the massless one.
The function 
$f_{2}(a,m_{0},d,\frac{u_{+}}{2})$ is non-singular in the bulk, i.e., in the 
interior of the interval $[0,a]$, while $f_{2}(a,m_{0},d,\frac{u_{-}}{2})$ has 
a singularity along the line $z=z^{\prime}$. Indeed, closer inspection shows that 
for $0\leq{}z,z^{\prime}\leq{}a$ the only singularities are those at $u_{+}=0$, $u_{+}=2a$ 
and also $u_{-}=0$. The former two are genuinely boundary singularities (the 
two conditions imply $z,z^{\prime}\rightarrow{}0$ or $z,z^{\prime}\rightarrow{}a$),
while the last comes from $z=z'$ in the whole domain and is just the standard 
bulk singularity. In fact, using the structure of the two-point 
function and showing just those terms from which singularities might arise, 
one finds that the counterterms for ${\tilde{M}}$ are given by
\begin{equation}
-\mbox{pole}\int_0^adz\,\int_0^adz^{\prime}[\frac{C_1}{(z+z^\prime)^{d-2}}+
\frac{C_2}{(2a-z-z^\prime)^{d-2}}
+\frac{C_3}{(z-z^\prime)^{d-2}}+...]^{2},
\end{equation}
where $C_{i}, i=1,..3$ are regular functions 
that do not depend on $z$ or $z'$. From this discussion it is clear 
that in order to render the field theory finite, 
we must introduce surface terms in the action. This is a 
general statement. For 
any fields that satisfy boundary conditions that break the translational invariance it is suffices to introduce
surface counterterms in the action, 
in addition to the usual bulk counterterms, to render the theory finite 
in the ultraviolet \cite{zy} \cite{di} \cite{ne}. Now we are able to 
discuss whether in the Casimir
configuration the infrared problems can be solved for the case 
of Neumann boundary conditions. For the case of massless 
$(\lambda\varphi^{4})_{d}$ theory at finite temperature, the infrared problem
can be solved after a resummation procedure \cite{linde} \cite{braaten} \cite{kapusta}  \cite{re2}  \cite{re3}. 
The key point for the solution of the infrared problem is to use the 
Dyson-Schwinger equation to rewrite the self-energy gap equation. Simple 
inspection of Eq.(\ref{19}) show us that it is not possible to implement
such scheme in a situation where there is a break of translational invariance.

A different possibility to approach the infrared problem is to 
single out the zero mode component of the field, treating the 
non-zero modes perturbativelly and treating the zero mode exactly.
This is a standard procedure in high-temperature field theory, where
by means of the dimensional reduction idea, we relate the thermal 
Schwinger functions in a d-dimensional Euclidean space to zero 
temperature Schwinger functions in a $(d-1)$ dimensional Euclidean space
\cite{jor} \cite{land} \cite{mar}. In this situation we have a dimensionally
reduced effective theory. The key point in this construction is the 
fact that the leading infrared behavior of 
any field theory at high temperature in a d-dimensional Euclidean space is governed by the zero frequency Matsubara mode.

\section{Discussions and conclusions}

In this paper we were interested in the analysis of the 
important questions of perturbative expansion 
and renormalization program in quantum field theory 
with boundary conditions that break translation symmetry, assuming 
that the system is in equilibrium with a reservoir at temperature
$\beta^{-1}$. Specifically, 
the purpose of this paper is to study the renormalization procedure up to one-loop level in 
the $(\lambda\varphi^{4})_{d}$ theory at finite temperature 
assuming that the scalar field satisfies 
Dirichlet-Dirichlet or  Neumann-Neumann boundary conditions 
on two parallel hyperplates.

We first obtained the regularized 
one-loop diagrams 
associated with scalar field defined in the Casimir 
configuration in a  d-dimensional Euclidean space. 
We obtained a well-know result concerning 
surface divergences 
that appear in the one-loop two-point and four-point Schwinger
functions as a consequence of the uncertaintly principle. There are
at least three different possible solutions to eliminate these 
divergences. The first one is to take into account that real materials have 
imperfect conductivity at high frequencies. 
As was stressed by many authors, the infinities that appear in renormalized 
values of local observables for the ideal conductor (or perfect mirror)
represent a breakdown of the perfect-conductor approximation. A wavelength 
cutoff corresponding to the finite plasma frequency must be included.  
The second one is is to substitute classical boundary conditions by classical potentials; for previous papers using this 
idea see for example \cite{mole} \cite{alb} \cite{hard}.
A localized boundary with some cut-off can also be used to replace the potential. Nevertheless, it is necessary to renormalize the potential \cite{caicedo}.
The third one regards a quantum mechanical treatment of the 
boundary conditions . A fruitful approach to avoid surface divergences, discussed by Kennedy et al. \cite{kea}
is to treat the boundary as a quantum mechanical object. This approach was developed by Ford and Svaiter \cite{ls} to produce finite values for the 
renormalized $<\varphi^{2}>$ and other quantities that diverge 
as one approach the classical boundary.

Consequently, we have two main distinct directions for future investigations.
The first is related to the infrared divergences of our model.  
Infrared divergences of massless thermal 
field theory arise from the zero frequency Matsubara modes, so 
we construct an effective $(d-1)$ dimensional theory
by integrating out the nonstatic modes and therefore the zero frequency Matsubara 
modes which are responsible for infrared divergences can be treated 
separately. 
The second direction is related to the surface divergences. 
In the Euclidean formalism for field theory,
one may imagine that our simplified model of rigid boundaries 
is a good approximation 
only for points in the bulk; for points close to the surfaces 
however, our
approximation is no longer acurate and a model taking into account 
at least thermal fluctuations of 
the boundaries must be developed \cite{kardar}. 
In other words, a fundamental understanding of the 
perturbative renormalization algorith in the standard 
weak-coupling perturbative expansion of an Euclidean field 
in the presence of fluctuating boundaries is desired.
This interesting situation of thermal fluctuating boundaries 
is under the investigation by the authors.

\section{Acknowledgement}

I would like to thanks L. A. Oliveira for comments on the manuscript.
This paper was supported in part by Conselho Nacional de
Desenvolvimento Cientifico e Tecnologico do Brazil 
(CNPq).

\end{document}